\documentclass[epj]{svjour}
\usepackage{graphicx}
\usepackage{amsmath}
\usepackage{amssymb}
\usepackage{amsfonts}
\begin{document}
\title{Incoherent Cooper pair tunneling and energy band dynamics in small Josephson junctions}
\subtitle{A study of the Bloch Oscillating Transistor}
\author{Ren\'{e} Lindell, Laura Korhonen, Antti Puska and Pertti Hakonen}
\authorrunning{R. Lindell, L. Korhonen, A. Puska and P. Hakonen}
\titlerunning{Incoherent Cooper pair tunnelling and energy band dynamics}

%
%
\institute{Low Temperature Laboratory, Helsinki University of
Technology, FIN-02015 TKK, Finland}
\date{Received: date / Revised version: date}
%
\abstract{ We discuss the properties of devices of small Josephson
junctions in the light of the phase fluctuation theory and the
energy band structure, which arises from the delocalization of the
phase variable. The theory is applied in the realization of a
mesoscopic amplifier, the Bloch oscillating transistor. The device
characteristics and comparison with theory and simulations are
discussed. The current gain of the device in a stable operating
mode has been measured to be as high as 30. Measurements on input
impedance and the power gain show that the BOT is an amplifier
designed for middle-range impedances, ranging from $100$
${\rm{k\Omega}} - 10$ ${\rm{M\Omega}}$.
} 
\maketitle

\section{Introduction}

The conjugate nature of charge and phase gives rise to a rich
array of physical phenomena, which have been intensively studied
both theoretically and experimentally. The quantum nature of the
phase variable was shown in macroscopic tunnelling experiments
\cite{clarke} and its conjugate relationship to the charge has
been shown in many consequent studies \cite{Tinkham1996}.

One of the consequences of the charge-phase conjugate relationship
is the Coulomb blockade of Cooper pairs which arises in very small
Josephson junctions \cite{haviland91}. The system is described in
phase or charge space, depending on which variable is a good
quantum number and, consequently, determines the eigenstates of
the quantum system. Associated with the two variables are two
competing energy scales, the charging energy $E_C = {\rm{e}}^2/2C$
and the Josephson coupling energy $E_J$. The Hamiltonian for the
small Josephson junction has a periodic potential, hence, Bloch
states and a corresponding energy band structure can be derived
$-$ analogously with the conduction electrons in solid state
physics \cite{Kittel,LZ}.

Incoherent tunnelling, the interaction of tunnelling electrons or
Cooper pairs with the electromagnetic environment, is a strong
effect observed in small tunnel junctions both in the normal and
superconducting states \cite{devoret90,holst94}. Small tunnel
junctions can be used as sensitive detectors of environmental
impedances and noise sources but, this sensitivity also means that
attention to control of the electromagnetic environment is
essential when designing mesoscopic devices.

The Bloch Oscillating transistor is a mesoscopic device which is
based on the dynamics of the Bloch bands in a voltage biased
Josephson junction (JJ) in a resistive environment. The main
operating principle was demonstrated in references
\cite{delahaye2,delahaye1,hassel04a,hassel04}. The conclusions that can be
drawn from these articles is that the BOT shows considerable
current and power gain (for both cases $\sim 30$). Additionally,
the noise temperature at the optimum operation point was observed
to be as low as 0.4 K, although, the theoretical and simulated
noise temperature has been shown to be 0.1 K or even below. The
BOT has an input impedance, which can fairly easily be tailored by
fabrication to be in the range $100$ ${\rm{k\Omega}} - 10$
${\rm{M\Omega}}$. Thus, the BOT is a device suitable for use at
mid-range impedances and it complements the group of basic
mesoscopic devices: the single-electron transistor (SET) - a
device for high-impedance applications and the SQUID, a low
impedance device.

At this point, we would like to emphasis that one should not
confuse the BOT device with the Bloch transistor \cite{AL}, which
is a similar but, fundamentally, a different kind of JJ device
where the current of a single Cooper pair transistor (SCPT) is
modulated by a gate voltage.

This paper will briefly review the physics that gives rise to the
voltage-current relation of the Josephson junction in light of
both the Bloch band model and incoherent Cooper pair tunnelling.
After the theory, a few observations on the "Bloch nose" - a
consequence of Bloch Oscillations - will be presented. The last
part of the paper will take on a more device oriented approach
with the discussion of the experiments and the question how well
the findings can be reproduced by theoretical simulations on the
Bloch Oscillating Transistor.
\section{Theory}
\subsection{Theory of Bloch states}

In mesoscopic systems the phase and charge comprise a conjugate
pair of quantum variables, analogously with the more familiar
space and momentum coordinates. The variables thus satisfy the
commutation relation
\begin{equation}
    \left [\phi,Q \right ] = 2ie.
    \label{commutation}
\end{equation}
The total classical Lagrangian for a single, isolated Josephson
junction can be written as
\begin{equation}
   \mathcal{L} = \frac{Q^2}{2C} + E_J \cos{\phi},
    \label{energy}
\end{equation}
which consists of the charging energy due to the capacitance $C$
and the Josephson coupling energy as the potential. From the
commutation relation (\ref{commutation}) we can immediately write
the quantum mechanical Hamiltonian \cite{LZ} as
\begin{equation}
    H = -E_C \frac{\partial^2}{\partial (\phi/2)^2} - E_J \cos{\phi}
    - \frac{\hslash}{2 {\rm{e}}} I \phi + H_{env} + H_{int},
    \label{Hamiltonian}
\end{equation}
where also the interaction between the driving current and the
phase variable $\propto -I\varphi$, the junction environment
energy $H_{env}$ and the coupling with the environment $H_{int}$
have been included. Depending on the theoretical framework used,
some, or none of the last three terms are included.

To set the initial stage, we consider only the first two terms in
the Hamiltonian, in the case when $E_C \gg E_J$ and the charge is
a good quantum number, thus leading to Coulomb blockade of Cooper
pairs and a complete delocalization of the phase. Equation
(\ref{Hamiltonian}) then takes the form of the familiar Mathieu
equation with the well-known solutions of the form
$\Psi_n^q(\varphi) = e^{i\varphi q/2e}u_n(\varphi)$, where
$u_n(\varphi)$ is a $2\pi$-periodic function and the wave
functions are indexed according to band number $n$ and quasicharge
$q$ \cite{quasicharge}. More precisely, the Mathieu equation has
also $4 \pi$ periodic solutions, which correspond to the the case
of quasiparticle tunnelling \cite{Schon}. The resulting energy
band structure is illustrated in Fig.~\ref{BOTschema}.
Verification of the existence of the energy bands has been carried
out by different methods \cite{prance92,flees97,lindell03}.

For the opposite limit, $E_J \gg E_C$, we should index the system
eigenstates according to the phase variable and the charge would
be completely delocalized. This situation corresponds to the
classical superconducting state of the Josephson junction, which
can also be described in the "tilted washboard" picture when
including the third term, $\propto I\phi$ in the Hamiltonian
(\ref{Hamiltonian}). This model can be easily extended to the well
known resistively and capacitively shunted junction (RCSJ) model,
where one includes a dissipative resistor in the circuit. But, we
will not discuss or apply the RCSJ model in this paper as it is
mostly used in the case of large Josephson junctions in the
context of macroscopic quantum tunnelling.

Next, we consider the Josephson junction in the band picture, when
the phase is delocalized and the state of the system can be
described by its quasicharge $q$. For a more realistic situation,
we also need to consider the junction's coupling to its
environment. The detailed and involved calculations for the cases
with ohmic and quasiparticle dissipation can be found in
Refs.~\cite{LZ,Schon}. The physics is rich in detail and includes
phenomena such as the zero-dimensional quantum phase transition
when the system goes from a superconducting $R < R_Q$ to an
insulating state $R > R_Q$ \cite{Penttila1999}.

\begin{figure}[bht]
\centering\includegraphics[width = 6.5cm]{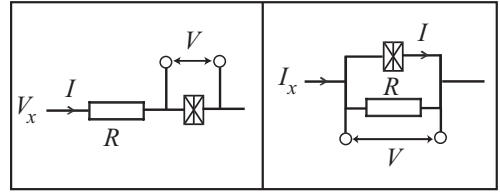}
\caption{Schematic view of the series or voltage bias (left) and
the parallel or current bias (right) configurations as defined in
the text. \label{jjcircuits}}
\end{figure}

The instantaneous voltage over the junction is given by
\begin{equation}
    V = \frac{{\rm{d}}E_n(q)}{{\rm{d}}q}.
    \label{avgvoltage}
\end{equation}
We consider the system to be either current or voltage biased,
meaning that the environmental resistor is situated in parallel or
in series with the junction (see Fig.~\ref{jjcircuits}). With a
steady current $I$ flowing through the junction the quasicharge
evolves according to
\begin{equation}
    \frac{{\rm{d}}q}{{\rm{d}}t} = I.
    \label{quasicharge}
\end{equation}

In the voltage biased case, the voltage over the total system
consists of the voltage drop over the series resistor and the
junction voltage
\begin{equation}
    V_x = IR + V.
    \label{voltbalance}
\end{equation}
In the current biased case, where the resistor is parallel with
the junction, we have the relation for the total current
\begin{equation}
    I_x = V/R + I.
    \label{currentbalance}
\end{equation}
Hence, one can switch between these two cases if one defines $I_x
= V_x/R$

In practice, we fabricate the resistor in series with the
junction, and thus consider next the former case of voltage bias.
If the resistor $R$ is large (compared to the CB resistance of the
junction), we can still think of the junction itself to be current
biased. Hence, if the driving current is low enough, ${\rm{d}}q /
{\rm{d}}t \ll e\delta E_1/\hslash$, where $\delta E_1$ is the gap
between the first and second band, the quasicharge $q$ is
increased adiabatically and the system stays in the first band. We
are then in the regime of Bloch oscillations; the voltage over the
junctions oscillates and Cooper pairs are tunnelling at the
borders of the Brillouin zone, or as for the definitions here (see
Fig.~\ref{BOTschema}), at $q = \pm e$. Consequently, the current
through the junction is coherent and the voltage and charge over
the junction oscillate with the frequency
\begin{equation}
    f_B = I/2e.
    \label{Blochfreq}
\end{equation}

The theoretical $I_x-V$ characteristics for an external current
bias thus first shows an increase of the junction voltage with
increasing current but at the onset of Bloch oscillations, the
voltage decreases. If the current $I$ is not adiabatically small,
we can have Zener tunnelling between adjacent energy bands. The
tunnelling is vertical, i.e the quasicharge does not change. The
probability of Zener tunnelling between bands $n-1$ and $n$ when
$E_C \gg E_J$ is given by
\begin{equation}
    P^Z_{n,n-1} = \exp \left (-\frac{\pi}{8} \frac{\delta E_n^2}{n
    E_C} \frac{{\rm{e}}}{\hslash I} \right ) = \exp \left (
    -\frac{I_Z}{I} \right ),
    \label{zener}
\end{equation}
where $\delta E_n = E_n - E_{n-1}$ and $I_Z$ is the Zener break
down current \cite{ben-jacob}.

The downward transitions can take place through several processes.
The cases we need to consider are transitions due to quasiparticle
tunnelling and due to charge fluctuations caused by the
environment. The quasiparticle transitions in the JJ couple $q$
states that differ by one $e$. In the BOT, the transitions are
primarily driven by a current bias through a second junction,
explicitly designed for returning the system to the lowest band.
The external environment gives rise to \textit{e.g.} current
fluctuations that couple linearly to the phase variable. These can
cause both upwards and downwards transitions. The strength of the
fluctuations is given by the size of the impedance: the larger the
impedance the smaller are the current fluctuations and the
transitions rates. As we will see later on, the successful
operation of the BOT requires one to control both the upwards and
downwards transition rates, or rather their relative strength.
Later, when discussing a very simple BOT model, we will make use
of the Zener transition rates and transitions due to charge
fluctuations, both derived in Ref.~\cite{Schon}.

\subsection{Incoherent tunneling and phase fluctuation theory}

It is well known that the electromagnetic environment of tunneling
junctions affects the tunneling process by allowing exchange of
energy between the two systems formed by the tunnel junction and
the external circuit \cite{devoret90,averin,girvin,IN}. The
influence of the external circuit can be taken into account, for
example, within the so called $P(E)$-theory, which is a
perturbation theory assuming weak tunneling. The external
environment is thought of as consisting of lumped element electric
components which give rise to a classical impedance $Z(\omega)$.
The impedance is quantized using the Caldeira-Leggett model, which
consists of an infinite number of LC-oscillators. Hence, in this
limit, one can model dissipative quantum mechanics. The energy fed
into the bath never returns to the tunnel junction system.

A perturbative treatment of the Josephson coupling term gives rise
to a simple looking result for incoherent Cooper pair tunneling
\cite{averin,IN}. The forward tunneling rate is directly
proportional to the probability of energy exchange with the
external environment:
\begin{equation}
    \overrightarrow{\Gamma}(V) = \frac{\pi}{2 \hslash} E_J^2
    P(2eV),
    \label{CPtunneling}
\end{equation}
and the backward tunneling rate is $ \overleftarrow{\Gamma}(V) =
 \overrightarrow{\Gamma}(-V)$, thus leading to the total current
\begin{equation}
    \begin{split}
    I(V) &= 2 {e} \left ( \overrightarrow{\Gamma}(V) - \overleftarrow{\Gamma}(V)  \right
    ) \\
    &= \frac{\pi {e} E_J^2}{\hslash} \left (P(2eV)  - P(-2eV)
    \right ).
    \end{split}
    \label{CPcurrent}
\end{equation}
The function $P(E)$ can be written as
\begin{equation}
    P(E) = \frac{1}{2 \pi \hslash} \int_{\infty}^{\infty}
    {\rm{d}}t \exp \left [J(t) + \frac{i}{\hslash} E t \right ],
    \label{PE}
\end{equation}
which is a Fourier transform of the exponential of the phase-phase
correlation function
\begin{equation}
    J(t) = \left < \left [\varphi(t)-\varphi(0) \right ] \varphi(0)
    \right>.
    \label{phasephase}
\end{equation}
The phase-phase correlation function is determined by the
fluctuations due to the environment and it can be related to the
environmental impedance with the fluctuation-dissipation theorem.
The result is that $J(t)$ can be found by

\begin{equation}
    \begin{split}
    J(t) = 2 \int_{0}^{\infty}
    \frac{{\rm{d}}\omega}{\omega}\frac{{\rm{Re}}  Z_t(\omega)}{R_Q}
    \{ & \coth(\beta \hslash \omega/2)
     [\cos(\omega t) - 1] \\
    & - i \sin(\omega t) \},
    \label{phasephase}
    \end{split}
\end{equation}
where $R_Q = h/4e^2$, and
\begin{equation}
Z_t(\omega) = \frac{1}{i\omega C + 1/R}
    \label{Zt}
\end{equation}
is the impedance seen by the junction.

We also need to consider quasiparticle tunneling, especially in
the BOT circuit, where we also have an NIS junction. We will not,
however, take into account Cooper pair transfer by Andreev
reflection \cite{Andreev1965} as it is not essential in the
experiment. For quasiparticles the perturbative tunneling rate
assumes the form

\begin{equation}
    \begin{split}
    \Gamma_{qp} & = \frac{1}{{\rm{e}}^2 R_T} \int_{-\infty}^{\infty} \int_{-\infty}^{\infty}
    {\rm{d}}E {\rm{d}}E' \frac{N_1(E)}{N_1(0)} \frac{N_2(E' +
    eV)}{N_2(0)}\\
    & \times f(E) [1-f(E'+eV)]P(E-E').
    \label{qprate}
    \end{split}
\end{equation}
Here, the density of states on the two sides of the junction
depend on whether the lead, $i = 1$ or 2, is normal or
superconducting. In the superconducting region, we have
$N_i(E)/N_i(0) = |E|/\sqrt{E^2 - \Delta^2}$, for $|E| > \Delta$
and zero otherwise, and in the normal region $N_i(E)/N_i(0) = 1$.

\subsection{BOT conceptual view}

\begin{figure}[bht]
\includegraphics[width = 8cm]{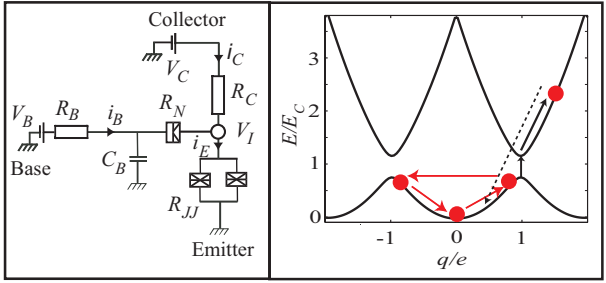}
\caption{Circuit layout of the BOT and conceptual view of its
operating cycle. The island is marked by $\bigcirc$. The two band
approximation of the BOT dynamics: Bloch oscillations (red
arrows), Zener tunneling and Coulomb blockade (black arrows) and
relaxation due to quasiparticle tunneling (black dashed arrow)
.\label{BOTschema}}
\end{figure}

The circuit schematics and a conceptual view of the BOT operation
cycle are shown in Fig.~\ref{BOTschema}. The basic circuit
elements are the Josephson junction, or SQUID, at the emitter,
with a total normal state tunnel resistance of $R_{JJ}$, the
single tunnel junction at the base with the normal state
resistance $R_N$, and the collector resistance $R_C$. The BOT base
is usually current biased with a large resistance $R_B$ at room
temperature, but the large stray capacitance $C_B$ results in an
effective voltage bias.  A requirement for the successful BOT operation
is that the charging energy $E_C = {\rm{e}}^2/(2C)$ is
of the same order of magnitude as the Josephson energy $E_J$. For
the theory presented in this paper to be valid, the condition
$E_J/E_C \ll 1$ has to be satisfied.

The theory describing the workings of the BOT is outlined in
articles \cite{hassel04a,delahaye2,hassel04}. The work by Hassel
and Sepp\"a \cite{hassel04}, contains both an analytical
approximation of the physics and the basic principles for a
simulation which are also used to some extent in this paper to
test the agreement between the experiment and simulation.


The basic physical principle of the BOT relies on the existence of
Bloch bands in the JJ, which is embedded in a resistive
environment $R_C \gg R_Q = h/4e^2$. The BOT is voltage biased to a
point on the lowest band where Bloch oscillations can start. The
supercurrent thus flows due to coherent Bloch oscillations (see
Fig.~\ref{BOTschema}) in the lowest band until the flow is stopped
by Zener tunneling into the second band. The JJ is Coulomb
blockaded until it relaxes down to the first band, either
intrinsically due to charge fluctuations caused by the
environmental resistance or with a controlling quasiparticle
current. The intrinsic relaxation is detrimental for BOT operation
and thus the fluctuations should be kept low by requiring that
$R_C \gg R_Q = h/4{\rm{e}}^2$. In practice, we need $R_C \gtrsim
100R_Q$ to be close to the ideal operation as outlined here. The
control current is injected through the second junction, which in
our case is a normal-insulator-superconductor (NIS) tunnel
junction. The amplification mechanism can, in its simplest form,
be said to arise from the number of Bloch oscillations triggered
by one quasiparticle. The Coulomb blockade of Cooper pairs (CBCP)
is thus a necessity for the BOT to function.

Our simulation of the BOT is based on the method given in
Ref.~\cite{hassel04}. In the simulation it is assumed that the
current flowing in the different parts of the BOT: the JJ, NIS
junction and resistor can be treated separately and the dynamics
of the island charge ($Q_I = V_IC_{JJ}$, where $C_{JJ}$ is the
capacitance of the Josephson junction and $V_I$ is the island
voltage indicated in Fig.~\ref{BOTschema}) is simulated and
averaged over a large number of steps (typically 10 - 100
million). The tunneling rates or, rather, the tunneling
probabilities in the JJ and NIS are calculated using the
$P(E)$-theory \cite{IN} described earlier. Furthermore, we assume
the same resistive environment for both junctions. The capacitance
in Eq.~\ref{Zt} is then $C = C_{JJ} + C_{NIS}$. The tunneling at
each point in time is then determined by comparing a random number
and the tunneling probability of the different junctions, which
depends on the voltage across them. Furthermore, the simulation
actually assumes that $E_J/E_C \ll 1$ and that the energy bands
can be approximated by parabolas: $E(q) = q^2/2C$. In this case,
the quasicharge is equal to the real island charge $Q_I$, which
is, according to the model in Fig.~\ref{BOTschema} also the charge
over the Josephson junction. The equation for the simulated island
charge is then given by

\begin{equation}
    \begin{split}
   \frac{{\rm{d}}Q_I}{{\rm{d}}t} = &\frac{V_C - V_I}{R_C} -
   \left(\frac{{\rm{d}}Q_I}{{\rm{d}}t}\right )_{QP_{NIS}} -
   \left(\frac{{\rm{d}}Q_I}{{\rm{d}}t}\right )_{QP_{JJ}} \\
   & -\left(\frac{{\rm{d}}Q_I}{{\rm{d}}t}\right )_{CP}.
   \end{split}
   \label{Islandcharge}
\end{equation}
Here the collector voltage $V_C$ takes the role of $V_x$ in the
theory of Bloch oscillations considered earlier and $V_I$ is the
island voltage. The island charge is thus modified by four terms:
the constant relaxation current through the collector resistor,
the quasiparticle tunneling through the base junction ($QP_{NIS}$)
and JJ ($QP_{JJ}$), and the Cooper pair tunneling in the JJ. The
theory thus excludes any quantum mechanical interactions between
the Josephson and NIS junctions. This simplification has shown to
be quite useful in determining the main operation principles of
the device. Although, one can argue that this simple treatment of
the island voltage as a time dependent variable while the
tunneling rates are unaffected by island dynamics is not the
correct way. As pointed out in Ref.~\cite{sonin}, the average
fluctuations of the phase of the island will govern the tunneling
in the Josephson junctions and, therefore, a more rigorous
approach would account for the peculiar phase fluctuations from
the tunneling quasiparticles. Nevertheless, time dependent
$P(E)$-theory should provide a reasonable starting point when the
environmental impedance is large \cite{soninPRL}. However, how to
include the Zener tunneling and the dynamics of the Bloch bands
into the phase fluctuation model is still under investigation.

\subsection{Alternative analytical theory}

The physical principle of the BOT current gain can also be derived
analytically from another viewpoint \cite{delahaye1}. The average
BOT emitter current can be thought of as the result of being in
either of the two states: the Bloch oscillation state with a
time-averaged constant current and the blockaded state with zero
current:

\begin{equation}
I_E = \left \{ \begin{aligned} V_C/R_C,& \qquad \tau_{\uparrow} = 1/\Gamma_{\uparrow}\\
0,& \qquad \tau_{\downarrow} = 1/(\Gamma_{\downarrow} + \Gamma_B).\\
\end{aligned}\right.
   \label{collectorcurrent}
\end{equation}
The amount of time the system spends in each state is given by the
Zener tunneling rate, $\Gamma_{\uparrow}$, the intrinsic
relaxation $\Gamma_{\downarrow}$, and the quasiparticle driven
relaxation $\Gamma_B$. The base current, however, flows during the
opposite times:
\begin{equation}
I_B = \left \{ \begin{aligned} 0,& \qquad \tau_{\uparrow} = 1/\Gamma_{\uparrow}\\
e \Gamma_B,& \qquad \tau_{\downarrow} = 1/(\Gamma_{\downarrow} + \Gamma_B).\\
\end{aligned}\right.
   \label{basecurrent}
\end{equation}
From these equations we can simply derive the average emitter and
base currents
\begin{equation}
\left < I_E \right > = \frac{V_C}{R_C}
\frac{\tau_{\uparrow}}{\tau_{\uparrow} + \tau_{\downarrow}}.
   \label{avgcollectorcurrent}
\end{equation}

\begin{equation}
\left < I_B \right > = -e \Gamma_B
\frac{\tau_{\downarrow}}{\tau_{\uparrow} + \tau_{\downarrow}}.
   \label{avgbasecurrent}
\end{equation}
From Eq.~(\ref{avgbasecurrent}) we can solve for $\Gamma_B$ and
insert this into Eq.~(\ref{avgcollectorcurrent}) in order to find
the emitter current
\begin{equation}
\left < I_E \right > = \frac{V_C}{R_C}
\frac{\Gamma_{\downarrow}}{\Gamma_{\uparrow} +
\Gamma_{\downarrow}}- \frac{V_C}{eR_C} \frac{1}{\Gamma_{\uparrow}
+ \Gamma_{\downarrow}} \left < I_B \right >.
   \label{avgcollectorcurrent2}
\end{equation}
We thus find the current gain
\begin{align}
    \beta_E = -\frac{\partial \left < I_E \right >}{\partial \left < I_B \right
    >} = \frac{V_C}{eR_C} \frac{1}{\Gamma_{\uparrow}
+ \Gamma_{\downarrow}}.
    \label{betac}
\end{align}
The base current relaxes through the collector resistance and,
therefore, the collector and emitter gains are related by
\begin{equation}
   \beta_E = \beta_C - 1.
   \label{betaec}
\end{equation}
The current gains are defined with the minus sign for convenience,
because, both theoretically and empirically, with the sign
convention used here the derivatives are negative.

Next, we have to find the transitions rates
$\Gamma_{\downarrow}(V_C)$ and $\Gamma_{\uparrow}(V_C)$ as a
function of the collector voltage. These have been calculated by
Zaikin and Golubev \cite{zaikin}. The Zener tunneling rate in a
resistive environment, and with the assumption $E_J \ll E_C$, is
given by
\begin{equation}
\Gamma_{\uparrow} = \frac{v}{2 \tau} \exp \left \{
-\frac{v_Z}{v-1}\left [ 1 + \frac{\left < \delta q^2/e^2 \right
>}{(v-1)^2} \right ]\right \},
   \label{uprate}
\end{equation}
and the down relaxation rate due to charge fluctuations is given
by
\begin{equation}
\Gamma_{\downarrow} = \frac{v_Z}{\tau \sqrt{2 \pi \left < \delta
q^2/e^2 \right >}} \exp \left \{ - \frac{(v-1)^2}{2 \left < \delta
q^2/e^2 \right >}{(v-1)^2} \right \},
   \label{downrate}
\end{equation}
where $v = CV_C/e$, $\tau = R_C C$  and
\begin{equation}
    v_Z =  \frac{\pi^2 R_C}{8 R_Q} \left ( \frac{E_J}{E_C} \right
    )^2.
   \label{zenertunneling}
\end{equation}
The voltage $v_Z$ is naturally related to the Zener break down
current $I_Z = ev_Z/(4 \tau)$. Equations (\ref{uprate}) and
(\ref{downrate}) are given as function of the bias voltage $v$,
although, the theory is strictly valid for the current bias case.
One can switch from the current biased to voltage biased case by
doing the transformation as discussed earlier by setting $I =
V_C/R_C$. In our experiment, the relevant case is the voltage
biased one.

Closed forms for the charge fluctuations in a resistive
environment can be found in the two limits of thermal and quantum
fluctuations:

\begin{equation*}
\left < \delta q^2 \right > = \left \{ \begin{aligned} k_B C T,& \qquad k_BT \gg \alpha_s E_C \\
\frac{2{\rm{e}}^2 \alpha_s}{\pi^2} \ln(\omega_c \tau),& \qquad k_BT \ll \alpha_s E_C,\\
\end{aligned}\right.
   \label{basecurrent}
\end{equation*}
where $\alpha_s = R_Q/R_C$. At the low temperature limit where
quantum fluctuations dominate one needs to know the cut-off
frequency $\omega_c$, which depends on the details of the external
impedance $Z(\omega)$. In practice, our experiments are done in a
regime where $k_BT \simeq \alpha_s E_C$, where both effects are
present.

The two-state model presented here has also been used for the
basic mechanism that generates the output noise of the BOT
\cite{hassel04}. The noise of the device is found to be output
governed and the input current noise is, therefore, inversely
proportional to the current gain \cite{lindell2005}.

\section{Fabrication and measurement}

\begin{table}[b]
\begin{tabular}{lllllllll}
\hline\noalign{\smallskip}
BOT \#&$R_{N}$&$R_{JJ}$&$R_C$& $E_J$&$E_C$&$\frac{E_J^{max}}{E_C}$&$\frac{E_J^{min}}{E_C}$\\
\noalign{\smallskip}\hline\noalign{\smallskip}
1 & 90 & 24  & 188 & 0.28 & 0.93 & 0.3 & 0.03\\
2 & 53 & 8.8  & 368 & 0.37 & 0.78 & 2.1 & 1.8\\
\noalign{\smallskip}\hline
\end{tabular}
 \caption{BOT parameters for two measured samples. $R_{N}$ and $R_{JJ}$ are the normal
state resistances of the NIS and JJ tunnel junctions,
respectively. Resistances are given in units of k$\rm{\Omega}$ and
energies in Kelvins. The tunability of the $E_J/E_C$ ratio is
indicated by the max and min values.} \label{BOTparams}
\end{table}

The fabrication of the BOT is done using standard electron beam
lithography and 4-angle shadow evaporation. The order of the
evaporation process was chromium (1), aluminum (2), oxidization
(3), chromium (4), copper (5) and aluminium (6). Originally, the
BOT device was thought to have an NIN junction as base junction,
but the technique of manufacturing both SIS and NIN junctions on
the same sample has not yet been mastered, therefore, we used NIS
junctions instead. In the first design, the NIS junction consisted
of the aluminium-aluminumoxide-copper interface, but it turned out
to be the most sensitive part of the device and it often was
already broken after bonding to sample holder. An additional, 7 nm
thick chromium layer, seemed to protect the junction and improved
the sample yield considerably.

Instead of having a single JJ at the emitter we made two
junctions, which thus form a SQUID loop. The SQUID configuration
gives the possibility of tuning the effective Josephson coupling
energy according to
\begin{equation}
   E_J = \sqrt{E_J^1 + E_J^2 + 2 E_J^1 E_J^2 \cos(\pi {\rm{\Phi / \Phi_0}})},
   \label{ejvsphi}
\end{equation}
where $E_J^1$ and $E_J^2$ are the Josephson energies for the two
junctions, $\rm{\Phi}$ is the externally applied magnetic flux
perpendicular to the loop area and ${\rm{\Phi_0}} = h/(2e)$ is the
quantum of flux. The $E_J$ tuning allows us to find the optimal
operation point in terms of the current gain \cite{delahaye2} as
well as giving the possibility to compare the $E_J$ dependence
with the theories presented in this paper. In practice, the
asymmetry of the SQUID junctions makes the tunability less than
perfect as can be noted from Table 1.

The BOT measurements were done on two different dilution
refrigerators: a plastic dilution refrigerator (PDR-50) from
Nanoway and the other from Leiden Cryogenics (MNK-126-500). Both
had a similar base temperature of 30 mK. The filtering in the PDR
consisted mainly of 70 cm long thermocoax cables on the sample
holder. Also, micro-wave filters from Mini-circuits (BLP 1.9) were
used at room-temperature. In the Leiden setup, besides the 1 m
long thermocoax on the sample holder, additional powder filters
(provided by Leiden Cryogenics) were present. Similar measurement
results were achieved with both setups, although the effective
temperature of the sample, deduced from the zero bias resistance,
was somewhat lower with the Leiden setup: 45-60 mK as compared to
80-100 mK with the Nanoway setup (see \textit{e.g.}
Ref.~\cite{lindellPRL}).

\begin{figure}[bht]
\center \includegraphics[width = 5.5cm]{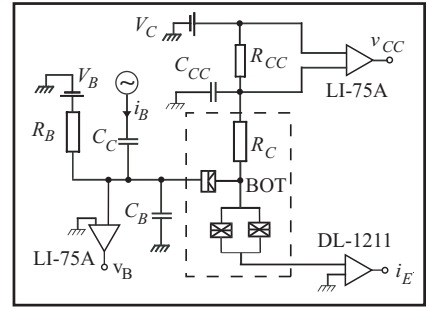}
\caption{Measurement scheme for power gain, current gain and input
impedance measurements. The BOT circuit is bounded by the red
dashed box. The AC signal is capacitively coupled through the
capacitor $C_C$. The signal from the pre-amplifiers LI-75A and DL
1211 are fed into lock-in amplifiers.
 \label{BOTschema2}}
\end{figure}

The measurement set-up used in this work is shown in
Fig.~\ref{BOTschema2}. Most of the complexity comes from the
requirements of the power gain measurement, which will be
presented in Sec. 4.2.4. The BOT base is DC current biased by a
large resistor $R_B$, which is located at room temperature. The
size of the resistor was typically $1-10$ G$\rm{\Omega}$. For
differential measurements, an AC-signal is fed through a coupling
capacitor $C_C$, in order to circumvent the large biasing
resistor. Voltages are measured with low noise LI-75A voltage
amplifiers and their output is fed into EG \& G Instruments 7260
DSP lock-in amplifiers. An additional, surface-mount resistor,
$R_{CC}$, is located on the sample holder, a few cm from the chip.
Therefore, this resistor is initially also at base temperature.
The role of the resistor is to act as load to the BOT, in order to
allow a clear measurement of power gain. The resistor was also in
used in noise measurements \cite{lindell2005} to convert current
noise to voltage noise.

\section{Experiment}
\subsection{Observation of the Bloch nose}

Our experiments correspond to the voltage biased configurations of
Fig.~\ref{jjcircuits}. The voltage is gradually lifted towards the
top of the energy band where it can overcome the first barrier and
start to oscillate. The $IV$ characteristics thus first shows an
increase of the junction voltage with zero current (when
neglecting any quasiparticle leakage) and then at the threshold
voltage where $V = \max \left \{{\rm{d}}E_n(q) / {\rm{d}}q \right
\}$ a current appears but the voltage over the junction goes down
as the time averaged voltage tends to zero. When the system
tunnels to a higher band, the Bloch oscillation is interrupted and
a voltage drop is once again formed over the junction. Thus, for
an increasing bias current, the voltage starts to rise again.

The actual measurement can be done with respect to the external
voltage $V_x$ (2-point measurement) or with respect to the real
junction voltage $V$: either by a 4-point measurement
\cite{watanabe2001,watanabe2003} or one can simply subtract the
known voltage $R_C I$ from the total (with the assumption that the
resistor $R_C$ is ohmic). According to the earlier transformation
between the serial and parallel configurations, we can compare the
$IV$ curves from our experiment with a series resistor with the
theoretical treatment in Ref.~\cite{Schon}, which is for the
parallel case by plotting $I_x = V_x/R_C$ against $V$ $-$ the
voltage over the JJ.

\begin{figure}[bht]
\includegraphics[width = 8.5cm]{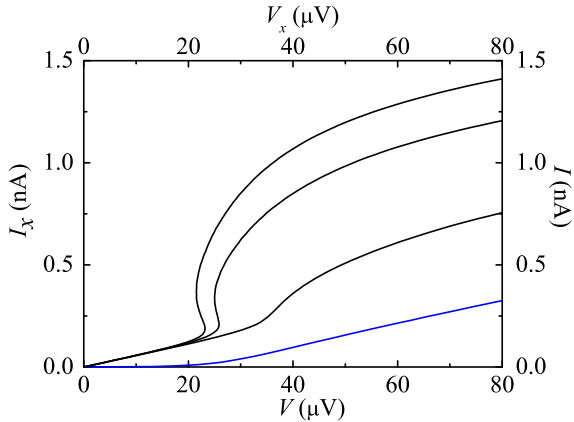}
\caption{$IV$ curve of sample \#1 from a 2-point measurement. The
lowest curve shows the total current $I$ vs.~$V_x$ (for $E_J/E_C =
$ 0.3) and the following curves are transformed to parallel
configuration $I_x = V_x/R_C$ vs.~ $V = V_x - IR_C$ with $E_J/E_C
= $ 0.3, 0.24 and 0.03 from top to bottom.\label{blochnoses}}
\end{figure}

In the experiments on the Bloch Oscillating Transistor, we have
also observed the Bloch nose and thus find the characteristic sign
of Bloch oscillation. The $IV$ curve of a Josephson junction in a
resistive environment for sample 1 (see parameters in Table 1) is
shown in Fig.~\ref{blochnoses}. The curves are a result of a
2-point measurement where both the total current $I$ and the
transformation $I_x = V_x/R_C$ are plotted against the total
voltage $V_x$ and the junction voltage $V = V_x - IR_C$,
respectively. In this way, one can truly see the Bloch nose in the
form predicted in Refs.~\cite{LZ,Schon}. The Bloch nose disappears
when the ratio $E_J/E_C$ goes down, a manifestation of the fact
that Zener tunneling sets in already at lower currents.
Theoretically, the voltage for the onset of back bending is given
by $V_b = 0.25e/C$, when $E_J/E_C \ll 1$ and the dissipation is
due to quasiparticles in the JJ \cite{Schon}. Without
quasiparticle tunneling the blockade would become much larger, and
for $E_J \ll E_C$ it would equal $e/C$. In our experiment, the
observed $V_b$ is 23 $\rm{\mu}$V, which is only little more than
half of theoretical value: $V_b = 0.25e/C = 40$ $\rm{\mu}$V.

The first report on the Coulomb blockade of Cooper pairs was made
by Haviland et al. in Ref.~\cite{haviland91}. Bloch oscillations
and the role of Zener tunneling was investigated by Kuzmin et al.
\cite{kuzmin94,kuzmin96} in experiments with a single Josephson
junction in an environment of a chromium resistor. The
characteristic Bloch nose feature, or the regions of negative
differential resistance, was first measured for 1- and
2-dimensional SQUID arrays \cite{haviland2000,geerligs1989}. In
later experiments using SQUID arrays as a tunable, but very
nonlinear, high impedance environment, the Bloch nose could for
the first time be clearly observed in a single Josephson junction
\cite{watanabe2001,watanabe2003}. Although, there the authors plot
$I$ vs.~$V$ and do not make the transformation $I_x = V_x/R_C$ as
done here. It should be noted, that our observation of the Bloch
nose involves a real, linear resistive environment as the
electromagnetic environment.

\subsection{Device characteristics}
\subsubsection{DC Current-voltage characteristics}

We will now describe the results of experiments of sample \#1 (
parameters are given in Table \ref{BOTparams}, whereas for the
definition of symbols, see Fig.~\ref{BOTschema2}) The maximum gain
was achieved for the maximum $E_J/E_C$ ratio (0.3 for this
sample). Hence, if not otherwise mentioned, all the following
presentations of the results are given for this value. According
to the theories on BOT operation
outlined earlier, the gain should increase with $E_J/E_C$.
However, the theories assume that $E_J \ll E_C$ so that it
suffices to consider only the dynamics between the first two
bands. In the BOT measurements described in Ref.~\cite{delahaye2}
the maximum gain was achieved for $E_J/E_C = 3.4$, which also
means that more than the two first bands may contribute to the
dynamics.

\begin{figure}[bht] 
\includegraphics[width = 8cm]{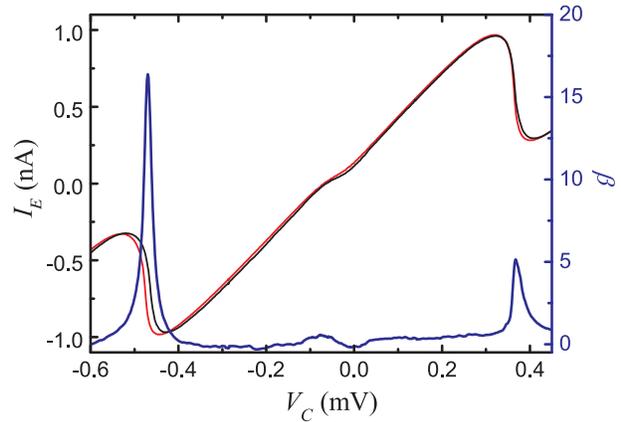}
\caption{IV curve for sample 1 at $I_B = 40$ pA (black curve) and
60 pA (red curve) and the DC-current gain (blue curve) calculated
by direct subtraction of the two IV curves. The extra $R_{CC} =
100$ k$\rm{\Omega}$ was located at collector. $T$ = 34 mK.
\label{ivcurve40}}
\end{figure}

\begin{figure}[bht] 
\includegraphics[width = 8.5cm]{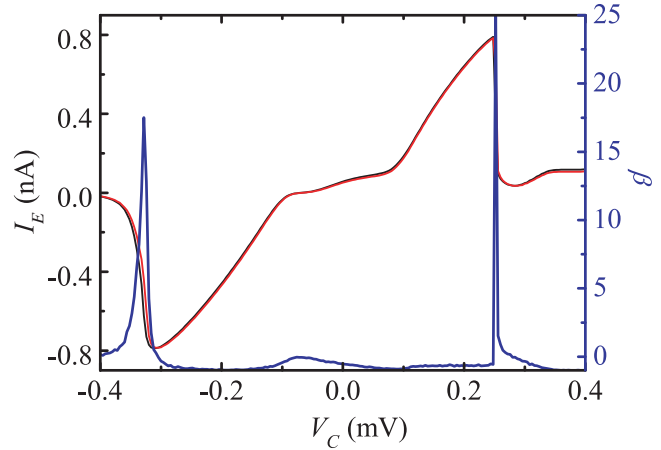}
\caption{Simulated IV curve with the same device parameters as in
the experiment presented in Figure \ref{ivcurve40}. Base current
100 pA (black curve) and 110 pA (red curve) and the DC-current
gain (blue curve) calculated by direct subtraction of the two IV
curves. \label{ivsimu100}}
\end{figure}

In Fig.~\ref{ivcurve40} a typical $IV$-curve of the BOT is shown
for two values of the base current and also the corresponding
current gain is included in the graph. The BOT has current
amplification for both positive and negative values of $V_C$.
However, as can be seen from the figure, the amplification is
largest for negative $V_C$. This we consider the "normal operating
mode" of the BOT and the gain region for positive $V_C$ we call the
"inverted mode". Simulated $IV$ curves and current gain are shown in
Fig.~\ref{ivsimu100} for the same device parameters as in the
experiment. A slightly larger base current than in the experiment,
100 pA as compared to 40 pA, was used to find a matching current
gain at the normal operating regime. The two sets of $IV$s' look quite similar. However, the
simulated one shows sharper features due to both a lower effective
temperature and a higher base current, which leads to hysteretic
behavior for the inverted operating region. Raising the
temperature or lowering the base current in the simulation leads
to a decreasing current gain. A prominent discrepancy between the
experimental and simulated $IV$'s is the location of the Bloch
"back bending" and the maximum gain. In the experiment, this
region is about 100 $\mu$V further than in the simulation. This is
an indication of the fact that the dynamics may actually involve
higher bands. More simulations for different BOT parameters can be
found in Ref.~\cite{hassel04}.

\subsubsection{Current gain}

\begin{figure}[bht]
\includegraphics[width = 8cm]{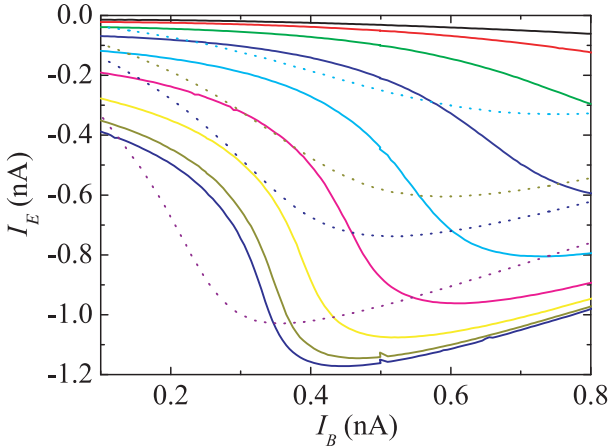}
\caption{$I_E$-$I_B$ curves for the normal region for different
$E_J/E_C$ ratios. The experimental curves (solid lines) go through
the ratios 0.03, 0.07, 0.12, 0.17, 0.21, 0.25, 0.28, 0.29 and 0.30
from top to bottom. The simulated curves (dotted lines) go through
the ratios 0.25, 0.28, 0.30 and 0.40 from top to bottom.
\label{icvsibnorm}}
\end{figure}

\begin{figure}[bht]
\includegraphics[width = 8cm]{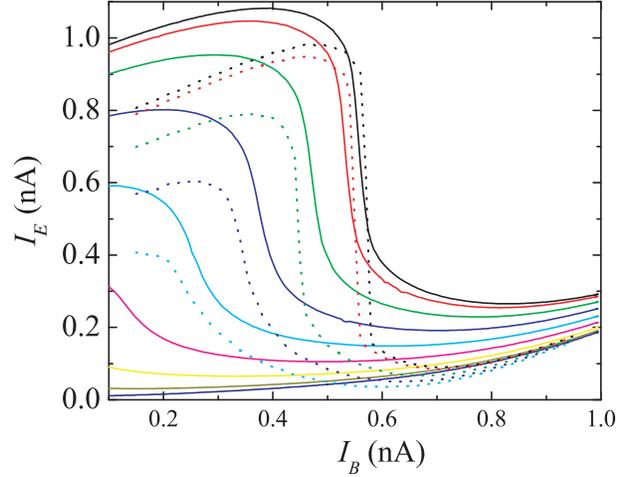}
\caption{$I_E$-$I_B$ curves for the region of inverted operation
for different $E_J/E_C$ ratios. The experimental curves (solid
lines) go through the ratios 0.03, 0.07, 0.12, 0.17, 0.21, 0.25,
0.28, 0.29 and 0.30 from bottom to up. The simulated curves
(dotted lines) go through the ratios 0.17, 0.21, 0.25, 0.29, 0.30
from bottom to top.\label{icvsibinv}}
\end{figure}

Next, we consider the $I_E - I_B$ relation for a fixed bias
voltage $V_C$ and as a function of the ratio $E_J/E_C$. In
Figs.~\ref{icvsibnorm} and \ref{icvsibinv} the $I_E - I_B$ curves
are shown for the normal and inverted operating regions. In these
measurements, the extra collector resistance $R_{CC}$ was absent,
hence we have larger operating currents. In the simulated $I_E -
I_B$ curves for the normal region, the gain was only about half of
that which was observed experimentally, and the gain region is
also much wider. However, by increasing the ratio $E_J/E_C$ in the
simulation to values slightly above those obtained from the
experiment (the simulated $E_J/E_C = 0.4$ is included in
Fig.~\ref{icvsibnorm}), the simulated gains become closer to the
experimental.

\begin{figure}[bht]
\includegraphics[width = 7cm]{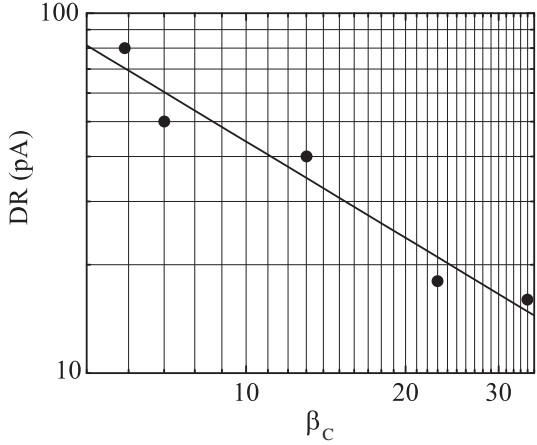}
\caption{Dynamic region (width at half maximum, in $I_B$-space) as
function of $\beta_C$ for the normal region for the base current
$I_B =$ 10, 20, 50, 70 and 100 pA (from left to right). The solid
line is a linear regression with slope  $-0.89 \pm 0.12$.
\label{DRvsbeta}}
\end{figure}

The simulation for the inverted mode showed diverging gains due to
hysteretic behavior that leads to very low dynamic region (see
also Fig.~\ref{DRvsbeta}). This approach to hysteretic and
divergent state could also be observed experimentally, as noticed
in Fig.~\ref{icvsibinv}.

\begin{figure}[bht]
\includegraphics[width = 8cm]{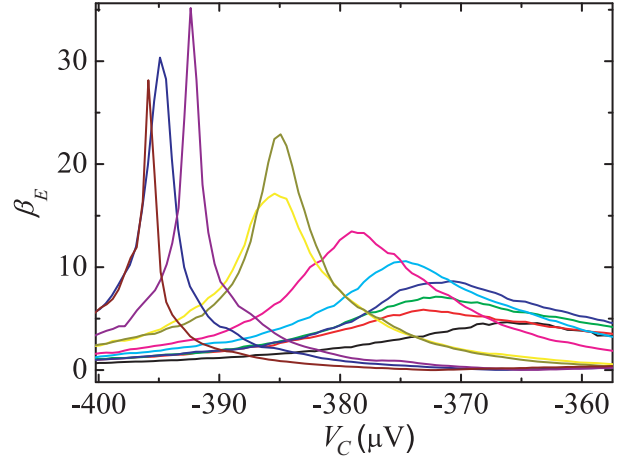}
\caption{The emitter current gain at the normal region for
$E_J/E_C = 0.3$ and the base currents 0-100 pA, with 10 pA
increments from right to left. The measurements were done using
the lock-in technique of Fig.~\ref{BOTschema2}. Here, the 100
k$\rm{\Omega}$ resistor at collector was present, thus leading to
smaller base currents for optimal operation. \label{gainsnorm}}
\end{figure}

We measured the differential gain as a function of $V_C$ with
lock-in amplifiers. The results for the normal operating region
are presented in Fig.~\ref{gainsnorm} for the case when the extra
collector resistance $R_{CC}$ was present. The gains are then a
factor of 3 larger than in the above measurement without $R_{CC}$.
This is also in close agreement to simulated gains, which will be
discussed further in section 4.2.3.

The dynamic region (DR) in terms of the base current $I_B$ can be
inferred from the gain curves in Fig.~\ref{gainsnorm} by using the
relation: $\Delta I_B = \Delta V_C/ (Z_{out} \beta_C)$. The
resulting DR versus $\beta_C$ plot is shown in Fig.~\ref{DRvsbeta}.
The large spread in the datapoints are mostly due to the errors in
determining $Z_{out}$ by numerical differentiation of $IV$ curves.
From the plot we see that the DR falls as $\propto \beta_C^{-1}$ (in
fact, the exponent of the regression is $-0.89 \pm 0.12$). As
already mentioned, the simulated gains for the normal region had a
much wider DR, though with less gain than in the experiment.

The analytic theory in Ref.~\cite{hassel04} predicts that the gain
should rise exponentially with the square of the ratio $E_J/E_C$:
\begin{equation}
    \beta_C = 2 \exp{\left [\frac{\pi {\rm{e}}^2 R_C}{8 \hslash}\left
    (\frac{E_J}{E_C} \right )^2 \right ]}.
   \label{zin}
\end{equation}

\begin{figure}[bht] 
\includegraphics[width = 8 cm]{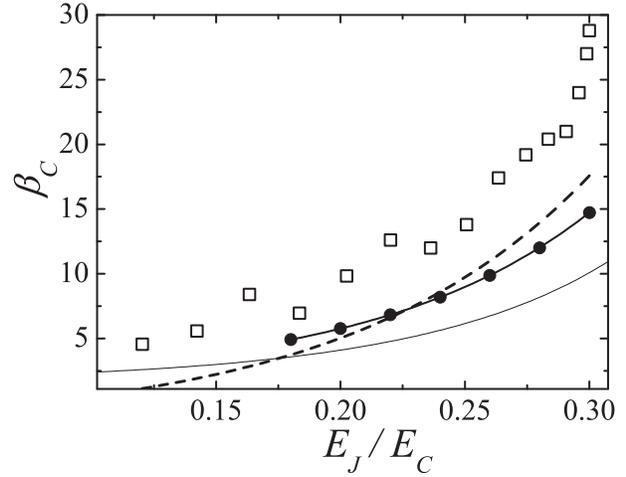}
\caption{Current gain $\beta_C$ as a function of $E_J/E_C$. The base
current $I_B$ was 100 pA both in the experiment and simulation.
Experiment normal operating mode ($\square$), simulation (dashed
line), the analytical model presented in this paper ($\bullet$)
and the analytical approximation from Ref.~\cite{hassel04} (thin
line). The temperature was assumed to be the same as base
temperature in the experiment, $T = 34$ mK.
 \label{betavsej}}
\end{figure}

According to our simple model presented in Sec.~$2.4$ the gain
dependence is non-monotonic. This could, however, be a failure of
the model as we are in a region were its validity is not
guaranteed by the perturbation theory. For $E_J/E_C = 0.1$ the
gain starts to rise again. From the full time-dependent
simulations, the dependence is also exponential but less than the
normal mode gains from experiment (Fig.~\ref{betavsej}). When
including heating effects in both the simulation and analytical
model, decreased the gain and made the fit worse.

\subsubsection{Input impedance}

\begin{figure}[bht]
\includegraphics[width = 8.0cm]{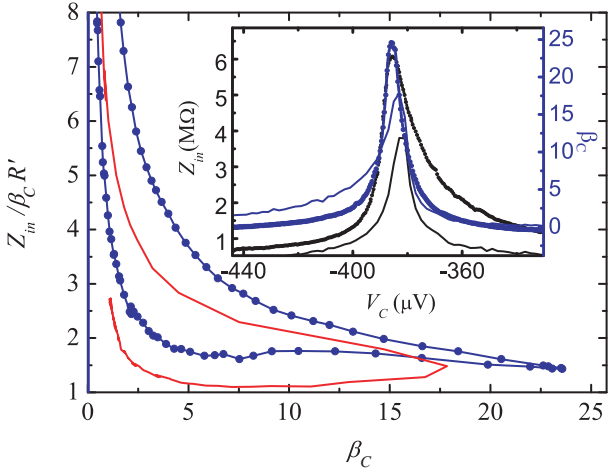}
\caption{$Z_{in}/(\beta_C R')$ as function of $\beta_C$ for $I_B = 60$
pA for experiment (blue curve) and simulated case (red curve). The
lower part of the curves is the behavior left of the maximum gain
and the upper part is for the right side. Inset: $Z_{in}$ and
$\beta_C$ as function of $V_C$ for experiment (thick lines) and
simulation (thin lines).
 \label{zinvsbeta}}
\end{figure}

The applications of the BOT are largely determined by its input
and output impedances, power gain and the noise temperature (for
measurements on the noise temperature see \cite{lindell2005}). We
have studied the input impedance
\begin{equation}
    Z_{in} = \frac{d v_B}{d i_B}.
   \label{zin}
\end{equation}
at different levels of the current gain. In Fig.~\ref{zinvsbeta}
the dependence is shown for the case where $I_B = 60$ pA. The
input impedance and $\beta_C$ were measured with lock-in amplifiers
with excitation frequency of 17.5 Hz. The measurement results had
to be corrected for the fact that part of the AC-current leaks
through the stray capacitance, which mostly originates from the
thermocoax. The corrected input impedance is then
\begin{equation}
    Z_{in} = \frac{z Z_C}{Z_C - z},
   \label{correctedzin}
\end{equation}
where $z = {\rm{d}}v'_B/{\rm{d}}i'_B$ is the measured
AC-differential impedance at the base and $Z_C = 1/(\omega C_B)$.
With the assumption that $C_B \simeq 600$ pF, the AC and DC gains
became equivalent.

The theoretical input impedance is easily calculated for the
"black box model" of Sec. 2.4, assuming lumped circuit elements
and a known, constant current gain $\beta_C$. Using the notation in
Fig.~\ref{BOTschema} we can write for the island voltage $V_I$
\begin{equation}
    V_I = V_C - R'I_C,
   \label{VI}
\end{equation}
where $R' = R_C + R_{CC}$, which includes the extra collector
resistor. On the other hand we also have
\begin{equation}
    V_I = V_B - I_B R_N.
   \label{VI2}
\end{equation}
As we know that $I_C = -\beta_C I_B$ we get from these two equations
\begin{equation}
    \Delta V_B = V_C + R'\beta_C I_B + I_B R_N,
   \label{VI2}
\end{equation}
and the input impedance is then
\begin{equation}
    Z_{in} = \frac{{\rm{d}}V_B}{{\rm{d}}I_B} = R_N + R'\beta_C.
   \label{zin}
\end{equation}
As $\beta_C$ is in our definition $>0$, the input impedance becomes
positive. The input impedance observed in the experiment
(Fig.~\ref{zinvsbeta}) shows some deviation from the simple model.
At the point of maximum gain the impedance is 1.4 the value of the
model, but as one moves to either side of the maximum, the
behavior is more complicated and also non-symmetric. The simulated
curve reveals that the behavior could be anticipated from the
dynamical model, which gives approximately the same 1.4 deviation
from the simple model as observed in the experiment. The
asymmetric feature shows that the simple model, that $Z_{in}$ is
proportional to $\beta_C$, is followed quite closely for the part of
the current gain where voltage is higher than the optimum
operating point.

\subsubsection{Power gain}

An important measure of the device performance is the power gain
\begin{equation}
\eta = P_{out}/P_{in},
   \label{eta}
\end{equation}
where $P_{in}  = i_B^2 Z_{in}$ and $P_{out}  = i_C^2 Z_{out}$, and
the currents are rms values. The simple black-box model for the
input impedance, where $Z_{in} \simeq \beta_C R_C$, gives a
theoretical power gain of
\begin{equation}
\eta = \frac{Z_{out}}{R_C}\beta_C.
   \label{etasimple}
\end{equation}

\begin{figure}[bht]
\includegraphics[width = 7cm]{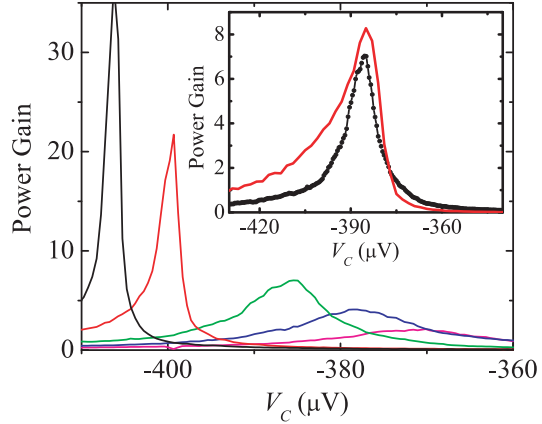}
\caption{Power gain with 100 k$\rm{\Omega}$ load for $I_B = $0,
20, 60, 80 and 100 pA from right to left and Inset: Power gain for
$I_B = 60$ pA ($\bullet$) and simulated curve (red
line)\label{pgain}}
\end{figure}

The setup  for the power gain measurement is shown in
Fig.~\ref{BOTschema2}. The AC-signal was capacitively coupled to
by-pass the large bias resistor $R_B$. We measured the power
delivered by the BOT to a 100 k$\rm{\Omega}$ resistor ($R_{CC}$),
which acted as a load at the collector. The measured power at the
output is given by $P_{out} = v_{CC}^2 /R_{CC}$, where $v_{CC}$ is
the measured AC-voltage over the load. The measured power gain in
this case becomes $\eta = (v_{CC}^2 / i_B^2)/ (R_{CC} Z_{in})$.
The simple black box model gives a power gain of $\eta =
(R_{CC}/R') \beta_C$. Taking into account the measured input
impedance in Fig.~\ref{zinvsbeta} we find that, for $I_B =  60$
pA, $\eta \simeq (R_{CC}/1.4R')\beta_C \simeq 6$, which agrees with
the independent measurement in Fig.~\ref{pgain}.

The power gains for $I_B = 0 - 100$ pA at the normal operating
point are shown in Fig.~\ref{pgain}. The largest measured power
gain was around 35 for $I_B = 100$ pA. The output impedance of the
device itself at the operating point was in the range $-50$
k$\rm{\Omega}$ to $-100$ k$\rm{\Omega}$. The gain might be
expected to grow rapidly when $R_{CC}$ approaches $|Z_{out}|$.
However, a positive load impedance with similar absolute magnitude
as the negative output impedance may lead to oscillatory behavior.

A simulated power gain for $I_B = 60$ pA is shown in the inset.
The agreement looks fairly good but when the base current is
increased in the simulation the effect on the gain is quite small
compared to the large increase observed in the experiment.
The simulation was also quite sensitive to the
capacitance $C_{CC}$ (see Fig.~\ref{BOTschema2}), here we settled
for a ratio $C_B/C_{CC} = 50$, which gave a good balance between
achieving a more realistic model and simulation stability. The
value of $C_B$ used in the simulation was 1 pF, which means that
the base voltage drops less than 0.1 \% when a quasiparticle
tunnels to the island.

\section{Conclusions}

We have here briefly reviewed the physical principles and
computational methods that have been applied to analyze the Bloch
Oscillating Transistor. The circuit has been shown to produce a
variety of interesting physical phenomena that also have a
potential for practical applications. The measurements on the BOT
have shown that the device works according to the physical
principles discussed. The Bloch nose as a manifestation of the
competition between coherent current via Bloch oscillations and
Coulomb blockade of Cooper pairs was observed in a controlled
resistive environment. The smallness of the observed blockade
would indicate that the environmental dissipation is dominated by
quasiparticles.

The optimal BOT parameters are quite difficult to specify, mainly,
because the used models are approximations that hold in a certain
regime for the parameters. We seek to optimize $R_N$, $R_{JJ}$,
$R$, $E_J/E_C$ and the capacitance $C_B$, and at the same time
keeping in mind what the optimum input and output impedances
should be for a specific application. The analytic model of Sec
2.4 gives a closed formula for the current gain, and can thus used
for optimization. In all the models, the gain grows with
$E_J/E_C$, but simultaneously, we have the requirement $E_J/E_C
\ll 1$ for the theories to be valid. However, nothing stops us to
go to somewhat larger $E_J/E_C$ ratios in the simulation,
although, the bands then increasingly deviate from the parabolic
approximation. The simulation shows that the gains also increase
with $E_J/E_C$, but naturally, with the expense of the dynamic
region. The simple analytic model does not depend on $R_N$ at all.
However, in the analytical model of Ref.~\cite{hassel04} the
device performance improves as $R_N$ is reduced. A too transparent
base junction would, however, be undesirable for many reasons: the
increased transparency could increase charge fluctuations and
destroy the Bloch oscillations in the Josephson junction. In fact,
in our samples the NIS junction was quite resistive. Also, small
$R_N$ would, in practice, mean larger capacitance and smaller
blockade. The model indicates that a larger collector resistance
$R_C$ is always desirable. This is probably a general conclusion,
bearing in mind that the input impedance increases linearly with
$R_C$. In practice, fabricating chromium on-chip resistors larger
than 0.5 M$\rm{\Omega}$ has proven difficult due to
nonlinearities.

The observed device properties of the BOT can be qualitatively
reproduced by simulations and the simple theoretical
considerations presented here. There are, however, still quite
many discrepancies between simulations and experiment. Often the
values of the simulation for gains and impedances were a factor 2
from the experimental case. The simulations do show that the main
operating principle of the BOT can be understood by the processes
outlined. There are still some places for improvement,
\textit{e.g.}, by taking into account the true phase fluctuations
caused by tunneling of quasiparticles to the island. But, this
would mean a great increase in the complexity of the simulation,
introducing new numerical challenges. Also, taking into account
the true band structure of the device might improve the agreement,
especially for $E_J > E_C$. While awaiting a more complete
theoretical treatment, the simulations can be used for qualitative
modelling of the BOT. The device could still be used as an on-chip
amplifier and detector for mesoscopic experiments, where
a good current gain is needed at medium level impedances, but, the
exact device properties can be inferred only from experiment.

\section*{Acknowledgments}
We acknowledge fruitful discussions with Julien Delahaye, Juha
Hassel, Frank Hekking, Mikko Paalanen and Heikki Sepp\"a.
Financial support by Academy of Finland, TEKES and Centennial
Foundation of Finnish Technology Industries is gratefully
acknowledged.

\section*{Appendix}

\subsubsection*{Transconductance}

The transconductance can be calculated in the simple black box
model by $g = \partial{I_C}/\partial{V_B} = -\beta_C
(\partial{I_B}/\partial{V_B}) = -\beta_C / (R_C \beta_C) = -1/R_C$.
For sample \#1, $1/R_C = 5.3$ $\mu$S and the maximum observed
transconductance was $5.0 \pm 0.5$ $\mu$S.

\begin{figure}[bht]
\includegraphics[width = 8cm]{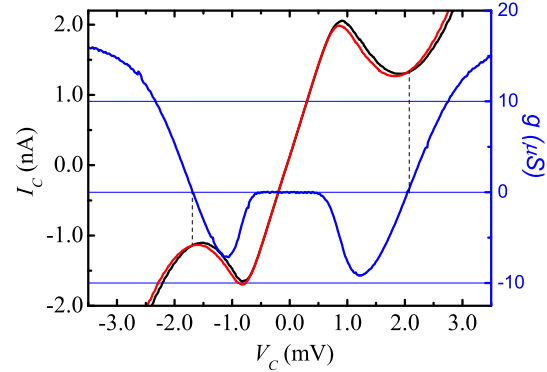}
\caption{$IV$ and transconductance curve for BOT \#2. $V_B = 0$
(black curve) and $V_B = 17.9$ $\mu$V (red curve). The dotted
lines show the points where the $IV$s cross and the
transconductance changes sign.\label{transc}}
\end{figure}

For sample \#2, which had a larger $R_C = 368$ k$\rm{\Omega}$, the
transconductance was 9.1 $\mu$S, which is about 3.5 larger than
expected from the simple model above (see Fig.~\ref{transc}). The
$IV$s cross at the dips in the $IV$-curves, thus giving rise to a
sign change in transconductance, and also for the current gain.
Thus, the input impedance $Z_{in} = \beta_C/g$ stays positive in all
measurements.

The $IV$ curve for sample \#2 was intrinsically hysteretic in
behavior when applying a current bias to the base via an
additional resistor at room-temperature. The hysteresis, however,
disappeared when the biasing resistor was between 100
k$\rm{\Omega}$ and 1 M$\rm{\Omega}$, which could indicate that the
device has a region of negative input impedance in this range.
Simulations have shown that the input impedance is always
positive, but that the base $I_B-V_B$ curve can be hysteretic and,
therefore, the observed hysteresis is a consequence of the device
switching between two operating regimes, which have the same base
current but different impedance levels.

%

%
%
%

\begin{thebibliography}{}
\bibitem{quasicharge}The quasicharge is analogous to the concept
of crystal momentum. Hence, the real charge is always given by
$\left < Q \right > = C \left < V \right > = C
\frac{\mathrm{d}E}{\mathrm{d}q}$.
\end{thebibliography}
%

\end{document}